\documentclass{iopart}

\usepackage{graphicx}
\usepackage{amssymb}

\begin{document}

\title[An Exactly Solvable Case for a Thin Elastic Rod]{An Exactly Solvable Case for a Thin Elastic Rod}

\author{Rossen Dandoloff$^\dag$ and Georgi G Grahovski$^{\dag , \ddag}$}

\address{$^\dag$Laboratoire de Physique Th\'eorique et Mod\'elisation,
Universit\'e de Cergy-Pontoise, 2 avenue A. Chauvin, F-95302
Cergy-Pontoise Cedex,
FRANCE\\
$^\ddag$Laboratory of Solitons, Coherence and Geometry, Institute
for Nuclear Research and Nuclear Energy, Bulgarian Academy of
Sciences,72 Tsarigradsko chauss\'ee, 1784 Sofia, BULGARIA}
\ead{rossen.dandoloff@u-cergy.fr $\qquad$  grah@inrne.bas.bg}

\begin{abstract}
We present a new exact solution for the twist of an asymmetric thin elastic rods. The shape of such rods is described by the static Kirchhoff equations. In the case  of constant curvatire and torsion the twist of the asymmetric rod represents a soliton lattice.
\end{abstract}


The study of elastic rods \cite{kirchhoff,love,ll7} has a long history and is a subject of
increased interest \cite{ng1,mg1,kdllt,al,Klapper} especially
in connection with bio-mathematical models of proteins and of DNA
\cite{lya,db1}. Thin elastic rods are described by the set of Kirchhoff equations \cite{kirchhoff}. On the other hand the static Kirchhoff rod model is also related to the dynamcs of spinning tops. In the prersent letter we will consider only the statics of thin elastic rods which is described by the static Kirchhoff equations.

Here we show that when the curvature and the torsion are constant, the twist of a thin
elastic asymmetric rod (i.e. with asymmetric cross section) can be
nonlinear along the rod. This represents a new sector of
integrable/ exactly solvable cases for the Kirchhoff rod.

Until now it has been assumed that the twist of an asymmetric
elastic rod is always  constant \cite{ll7}. Now it is clear that
the curvature of the rod plays an essential r\^ole and it may
``interact'' with the asymmetry of the cross section in order to
produce a new class of highly nontrivial (soliton lattice) solutions.

The Kirchhoff rod represents in general a non-integrable  system
\cite{love,ng1,kdllt}. Nevertheless there are special cases which
are integrable:  where the twist along the rod and the torsion
along the cental line of the rod are constant.
The case when the
torsion and the twist are constant along the rod is widely
discussed in the literature \cite{ll7}.

A thin rod can be modelled by a space curve ${\bf R}(s)$ joining
the loci of the centroids of the cross sections together with the
(so-called) Frenet frame $({\bf t}(s),{\bf n}(s),{\bf b}(s))$
attached to the rod material and evolving along the shape
according to the Frenet--Serret equations (with an arc length
parameter $s$)
\begin{eqnarray}\label{eq:fren-ser}
{d{\bf t}\over ds}=\kappa {\bf n}(s),\quad {d{\bf n}\over
ds}=-\kappa {\bf t}(s)+\tau {\bf b}(s),\quad {d{\bf b}\over
ds}=-\tau {\bf n}(s)
\end{eqnarray}
Here  ${\bf t}_i(s)$, ${\bf n}_i(s)$ and ${\bf b}_i(s)$ are the
tangent, the normal, and the binormal   vectors to the curve
respectively. The tangent vector is a unit vector given by ${\bf
t}=\left({ d{\bf R}\over ds}\right) $, the curvature $\kappa (s) $
of the curve at the point $s$ is then given by: $ \kappa (s):=
\left|{d{\bf t}\over ds}\right|$ and the normal and the binormal
vectors can be determined from (\ref{eq:fren-ser}).  The local
basis $({\bf d}_1(s),{\bf d}_2(s),{\bf d}_2(s))$ which is attached
to the rod can be expressed through the Frenet frame as follows:
\begin{eqnarray}\label{eq:localbas}
{\bf d}_1(s) &=&-{\bf n}(s)\sin \phi + {\bf b}(s)\cos \phi \\
{\bf d}_2(s)&=&{\bf n}(s)\cos \phi + {\bf b}(s)\sin \phi,\qquad  {\bf d}_3(s) ={\bf t}(s) \nonumber
\end{eqnarray}
where $\phi $ is the twist angle of the rod. The components of the
derivatives of the local basis $( {\bf d}_3(s), {\bf d}_2(s), {\bf
d}_1(s))$ with respect to $s$ can be expressed by using the twist
vector ${\bf k}(s)=\kappa_1{\bf d}_1+\kappa_2{\bf d}_2 +
\kappa_3{\bf d}_3$:
\[
{d {\bf d}_i\over ds} = {\bf k}(s) \times {\bf d}_i(s), \qquad i=1,2,3.
\]

The static Kirchhoff equations describe the shape of the rod
under the effects of internal elastic stresses and boundary
constraints, in the absence of external force fields. If ${\bf
F}(s)$ denotes the tension and ${\bf M}(s)$ denotes the torque of the rod, then
in the generic case the torque is
\begin{eqnarray}\label{eq:torque}
{\bf M}(s)= \kappa_1(s){\bf d}_1(s) + a\kappa_2(s){\bf d}_2(s) + b\kappa_3(s){\bf d}_3(s)
\end{eqnarray}
The constant $a$ measures the asymmetry of the cross section and
$b$ is the scaled torsional stiffness. In particular for symmetric
($a=1$) hyperelastic ($b=1$) rods  we have ${\bf M}(s)={\bf
k}(s)$. The elastic energy of the Kirchhoff rod is given by:
\begin{eqnarray}\label{eq:Helast}
H= {1\over 2} \int_{0}^{s}{\bf M}\cdot {\bf k}\,ds= {1\over 2}
\int_{0}^{s}\left( \kappa_1^2 + a\kappa_2^2 +
b\kappa_3^2\right)\,ds.
\end{eqnarray}
The conservation of the linear and
angular momenta is provided by the static Kirchhoff equations:
\begin{eqnarray}\label{eq:Kirchhoff-m}
{d{\bf F}\over ds}=0, \qquad {d{\bf M}\over ds}+ {\bf
d}_3(s)\times {\bf F}(s)=0.
\end{eqnarray}
The twist vector ${\bf k}$ is related to the curvature $\kappa
(s)$ and to  the torsion $\tau (s)$ of the center line of the rod
by:
\begin{eqnarray}\label{eq:kappa_0}
{\bf k}(s)= (k_1,k_2,k_3)=(\kappa \sin \phi , \kappa \cos \phi , \tau + \phi_s ),
\end{eqnarray}
where the twist angle $\phi $ is a function of the arc length
parameter $s$: $\phi=\phi (s)$.  Then the vector equations
(\ref{eq:Kirchhoff-m}),  projected onto the local basis $({\bf d}_1,
{\bf d}_2, {\bf d}_3)$ give the following 6 scalar equations for
the tension ${\bf F}=(F_1,F_2,F_3)$ and the twist ${\bf
k}=(k_1,k_2,k_3)$:
\begin{eqnarray}\label{eq:Krichhoff-helix-f1s}
&&{dF_{1}\over ds}+ \kappa_2 F_3 -\kappa_3 F_2=0 \\\label{eq:Krichhoff-helix-f2s}
&&{dF_{2}\over ds}+ \kappa_3 F_1 -\kappa_1 F_3=0 \\\label{eq:Krichhoff-helix-f3s}
&&{dF_{3}\over ds}+ \kappa_1 F_2 -\kappa_2 F_1=0 \\\label{eq:Krichhoff-helix-f1}
&&F_1=-a{d\kappa_{2}\over ds}+ (b-1)\kappa_1\kappa_3\\\label{eq:Krichhoff-helix-f2}
&&F_2={d\kappa_{1}\over ds}+ (b-a)\kappa_2\kappa_3\\\label{eq:Krichhoff-helix-f3}
&&b{d\kappa_{3}\over ds}+(a-1)\kappa_1\kappa_2=0.
\end{eqnarray}
In this system of equations we replace $k_1$, $k_2$ and $k_3$ from eqn. (\ref{eq:kappa_0}).

In the case of constant curvature $k(s)=k_0$ and  torsion $\tau
(s)=\tau_0$ the equation (\ref{eq:Krichhoff-helix-f3}) reduces to
the famous static (scalar) sine--Gordon equation for the twist
angle:
\begin{eqnarray}\label{eq:Krichhoff-helix-sg}
{d^2 u \over ds^2} + {(a-1) \over b} \kappa_0^2 \sin u(s)=0, \qquad u(s)=2\phi(s).
\end{eqnarray}
This second order differential equation represents in itself a completely integrable
Hamiltonian  system and allows so-called ``soliton''-like
solutions. It appears in a wide variety of physical problems for
e.g. charge-density-wave materials, splay waves in membranes,
magnetic flux in Josephson lines, torsion coupled pendula,
propagation of crystal dislocations, Bloch wall motion in magnetic
crystals, two-dimensional elementary particle models in the
quantum field theory, etc.

The nontrivial solutions of eq. (\ref{eq:Krichhoff-helix-sg}) are single solitons or kinks and soliton lattices. We will show that the kink-solutions of (\ref{eq:Krichhoff-helix-sg}) are not compatible with the full set of Kirchhoff equations.

In what follows we will prove that the periodic (soliton-lattice) solution of
(\ref{eq:Krichhoff-helix-sg}) is compatible with the full system (\ref{eq:Krichhoff-helix-f1s}) - (\ref{eq:Krichhoff-helix-f3}) and thus represents an exact solution for the thin rod in the case of constant curvature and torsion. This solution is given by the following expression \cite{skyrme}:
\begin{eqnarray}\label{eq:periodic}
\phi(s)=2\, \mbox{arccos}\, \left[ \mbox{sn}\, \left(
{\kappa_0\over m}\sqrt{{1-a\over b}}s, m\right)\right],
\end{eqnarray}
where $m$ is the modulus of the Jacobian elliptic function $sn$, and $K(m)$ is the complete elliptic integral of
the first kind. In the limit $m \to 1$  $K(m)\to \infty$
and the period tends to infinity. The period of the solution is
\[
d=\sqrt{{b\over 1-a}}{1\over k_0}4mK(m).
\]
In order to prove the compatibility of (\ref{eq:Krichhoff-helix-sg})  with
the full system of Kirchhoff  equations
(\ref{eq:Krichhoff-helix-f1s})--(\ref{eq:Krichhoff-helix-f3}) we follow the ideas of \cite{gnt1}. Let us consider the following linear combination of (\ref{eq:Krichhoff-helix-f1s}) and (\ref{eq:Krichhoff-helix-f2s}) taking into account the explicit parametrization (\ref{eq:kappa_0}) of the twist vector ${\bf k}$:
\begin{eqnarray}\label{eq:comb-1}
\fl [F_{1,s}-(\tau_0 + \phi_s)F_2 + k_0F_3 \cos \phi]\sin \phi + [F_{2,s}+(\tau_0 + \phi_s)F_2 - k_0F_3 \sin \phi]\cos \phi=0,
\end{eqnarray}
which reduces to
\begin{eqnarray}\label{eq:comb-2}
F_{1,s}\sin \phi + F_{2,s}\cos \phi + (\tau_0 + \phi_s)[F_{1}\cos \phi - F_{2}\sin \phi]=0.
\end{eqnarray}
Here and below the subscript $s$ means  a derivative with respect to $s$. Substituting (\ref{eq:Krichhoff-helix-f2}) and (\ref{eq:Krichhoff-helix-f3}) into (\ref{eq:comb-2}) yields
\begin{eqnarray}\label{eq:comb-3}
 [4b + 2(1-a)\cos 2 \phi]\phi_{ss}=(1-a)(2\phi_s +\tau_0)^2\sin 2\phi,
\end{eqnarray}
for the case of nonvanishing curvature $k$. Then we use the expression for $\phi_{ss}$ from (\ref{eq:Krichhoff-helix-f3}) and write the above equation in the following form
\begin{eqnarray}\label{eq:comb-4}
(1-a)k_0^2 \sin 4 \phi + 2b[2k_0^2-4\phi_s^2 - 4\tau_0\phi_s - \tau_0^2]\sin 2\phi=0.
\end{eqnarray}
After multiplication of both sides of (\ref{eq:Krichhoff-helix-f3}) by $\phi_s$ and integration we get:
\begin{eqnarray}\label{eq:phi_s}
b\phi_s^2 =(1-a)k_0^2\sin^2 \phi + C_0,
\end{eqnarray}
where $C_0$ is an integration constant. Now we eliminate $\phi_s^2$ from (\ref{eq:comb-4}) and (\ref{eq:phi_s}) and get
\begin{eqnarray}\label{eq:comb-5}
4b\tau_0\phi_s=3(1-a)k_0^2 \cos 2 \phi + 2(b+a-1)k_0^2-4C_0 - b\tau_0^2.
\end{eqnarray}
Next we square both sides of the last equation and eliminate $\phi_s^2$ using eq. (\ref{eq:phi_s}). Then eqn. (\ref{eq:comb-5}) transforms into a polinomial for $\cos 2\phi$:
\begin{eqnarray}\label{eq:comb-poll}
A\cos^2 2\phi + B \cos 2\phi + C =0,
\end{eqnarray}
The coefficients take the form
\begin{eqnarray}\label{eq:comb-coeff}
A&=&-9(1-a)^2k_0^4,\\
B&=&-8b\tau_0^2(1-a)k_0^2-6(1-a)k_0^2[2(b+a-1)k_0^2-4C_0-b\tau_0^2],\\
C&=& 8b\tau_0^2[(1-a)k_0^2+2C_0] - [2(b+a-1)k_0^2-4C_0-b\tau_0^2]^2.
\end{eqnarray}
If we replace in eqn. (\ref{eq:comb-poll})  $\cos 2\phi$ with $\sin 2\phi$ and $\cos \phi$ with $\sin \phi$ we will cast the trigonometric polynomial in the following form:
\begin{eqnarray}\label{eq:comb-poll2}
-A\sin^2 2\phi -2B \sin^2 \phi +A+B+C=0.
\end{eqnarray}
The above expression must vanish identically for $\phi (s)$  from eq. (\ref{eq:periodic})if this solution  is to be compatible with the full set of Kirchhoff equations. First let us note that $\phi$ from eqn. (\ref{eq:periodic}) satisfies the following relation \cite{prl}
\begin{eqnarray}\label{eq:dens}
\phi_s^2={1-a\over b}k_0^2\sin^2 \phi + C_0,
\end{eqnarray}
where $C_0={1-a\over b}k_0^2\left( {m^\prime \over m}\right)^2$.
Now we use eq. (\ref{eq:dens}) to replace $\sin^2 \phi$ and
$\sin^2 2\phi$ in eqn. (\ref{eq:comb-poll2}). The expression in
(\ref{eq:comb-poll2}) vanishes identically if the following two
conditions are fulfilled:
\begin{eqnarray}\label{eq:compat-cond}
 B=-2A,\qquad  A+B+C= 3\left( {m^\prime\over m}\right)^2.
\end{eqnarray}
Inserting the first into the second equation leads to:
\begin{eqnarray}\label{eq:compat-cond2}
C=A\left[ 1+3\left( {m^\prime\over m}\right)^2\right].
\end{eqnarray}
These two conditions are easily fulfilled. The first equation leads to a condition on the integration constant $C_0 $ and the second one with the use of the first gives an algebraic equation of fourth order for the curvature $k_0$
Solving eqn.(\ref{eq:compat-cond}) for $C_0$ gives:
\begin{eqnarray}\label{eq:compat-cond3}
C_0={1-a\over b}k_0^2\left( {m^\prime\over m}\right)^2= {b\tau_0^2\over 12} + \left[ {1-a\over 4} + {b\over 2}\right]k_0^2.
\end{eqnarray}
We note here that $C_0$ must be positive for the nontrivial case $k_0\neq 0$ and $\tau_0\neq 0$.
Finally the condition (\ref{eq:compat-cond2}) leads to the following equation:
\[
\fl {27\over 4}(b-a)(2b-a+1)k_0^4 + \tau_0^2\left[ 4b(2b-a+1)+{9\over 4}b^2(1-a)\right] k_0^2-{1\over 3}b^2\tau_0^4=0.
\]
This equation has always positive solutions for $k_0^2$ and thus we have shown that $\phi(s)$ from (\ref{eq:periodic}) is compatible with the full set of static Kirchhoff equations.

As we noted before eqn. (\ref{eq:Krichhoff-helix-sg}) has also a
kink solution $\phi (s)=\arctan {\rm e}^{s/s_0}$ with $s_0={b\over
(1-a)k_0}$. For this solution it is easy to show  that the
constant $C_0$ in (\ref{eq:dens}) must be zero. This is only
possible if $k_0=0$ and $\tau_0=0$ (see eq.
(\ref{eq:compat-cond3})). In this case however the equation
(\ref{eq:Krichhoff-helix-f3}) turns into a linear ODE for
$\phi(s)$: $\phi_{ss}=0$.

Now we turn our attention to the elastic energy density for the thin rod. We will show that the new exact nonlinear solution for the thin elastic rod may be energetically more favorable than the trivial no twist solution $\phi=0$.

The energy density (per unit length) of the rod is:
\begin{eqnarray}\label{helix1}
h=\frac{1}{2}[ b(\tau_0+\phi_s)^2+(1-a)k_0^2\sin^2\phi+ak_0^2]
\end{eqnarray}
The Euler-Lagrange equation for the energy is once again the
eqn.(13) and its nontrivial solution is given in eqn.(14) with
$s_0=\frac{1}{k_0}\sqrt{\frac{b}{1-a}}$.
The corresponding static energy per soliton of the soliton lattice
is given by:
\begin{eqnarray}\label{eq:E-periodic}
\fl E_{\rm twist}= l(b\tau_0^2+ak_0^2)+ 2\tau_0[\phi(l)-\phi(-l)]+
{\kappa_0\over m}\sqrt{{1-a\over b}} \left(E(m)-{1\over 3}(m^{\prime})^2K(m) \right),
\end{eqnarray}
where $E(m)$ is the complete elliptic integral of second kind and $m^{\prime}$ is the complementary modulus: $m^{\prime}=\sqrt{1-m^2}$.

Now let us compare this energy $E_{\rm twist}$ with the energy of
the trivial solution: If there is no twist  $(\phi = 0, \pi, 2\pi,
...)$ the energy for length $l$ which corresponds to one period
$d$ of the soliton lattice
\begin{eqnarray}\label{helix3}
E_{\rm no\, twist}=l(b\tau_0^2+ak_0^2)
\end{eqnarray}
For some choice of the parameters $k_0,\tau_0,a,b$, $E_{\rm
twist}$ may turn smaller than $E_{\rm no \, twist}$. This may
happen if the following condition holds true (depending on the
choice of $\tau_0$, $\kappa_0$, $a$, $b$ and $m$):
\[
\fl 2\tau_0[\phi(l)-\phi(-l)]+
{\kappa_0\over m}\sqrt{{1-a\over b}} \left(E(m)-{1\over 3}(m^{\prime})^2K(m) \right) \leq 4\pi\tau_0-{1\over 3}
{\kappa_0^2\over m}{1-a\over b}d <0,
\]
where $d$ is the period of the soliton lattice. We have used  that
$|\phi(l)-\phi(-l)|\leq 2\pi$. In this case the static energy per
unit length of the solution lattice will be lower than the static
energy in the trivial no twist case.

With this new exactly solvable case for the thin elastic rod the following general picture for the twist of thin rods emerges:

(A) The twist for all symmetric thin rods ($a=1$) with constant torsion $\tau_0=\mbox{const}$ is either constant $\phi(s)=\mbox{const}$ or periodic and {\it linear} in $s$: $\phi(s)=\mbox{const}. s$. Here the twist is disconnected from the curvature.

(B) The twist for all asymmetric thin rods ($a\neq 1$) (with constant curvature $\kappa_0$ and torsion $\tau_0$) is either constant: $\phi =n\pi/2$ or periodic and {\it nonlinear} in $s$: $\phi(s)=2\arccos [\mbox{sn} (s/s_0,m)]$. Here the asymmetry $(1-a)$ couples the twist with the curvature $k_0$ which leads to this nonlinear solution.

\section*{Acknowledgments}

We would like to thank V. S. Gerdjikov, N.
A. Kostov and Radha Balakrishnan for many valuable discussions.
The work of GGG is supported by the Bulgarian National Scientific
Foundation Young Scientists Scholarship for the project
``Solitons, Differential Geometry and Biophysical Models''.The
support by the National Science Foundation of Bulgaria, contract
No. F-1410 is also acknowledged.


\bigskip

\begin{thebibliography}{99}

\bibitem{kirchhoff} G. Kirchhoff 1859 J Reine Angew. Math {\bf 56}  285.

\bibitem{love} A. E H. Love 1944 {\it A Treatise on the Mathematical Theory od Elasticity}, Dover Publications, New York.

\bibitem{ll7} L. D. Landau and E. M. Lifshitz 1986 {\it Theory of Elasticity} (Course of Theoretical Physics, Vol 7), Pergamon Press, Oxford.

\bibitem{ng1} M. Nizette and A. Goriely 1999   J. Math. Phys. {\bf 40} 2830--2866.

\bibitem{mg1} T. McMillen and A. Goriely 2002   J. Nonlin. Sci. {\bf 12}  241--281.

\bibitem{kdllt} B. D. Coleman, E. H. Dill, M. Lembo, Zh. Lu and I. Tobias 1993 Arch. Rational Mech. Anal. {\bf 121} 339--359.

\bibitem{al} S. S. Antman and  T.-P.Liu 1978/79  Quart. Appl. Math.  {\bf 36}   no. 4 377--399.

\bibitem{Klapper} I. Klapper 1996  J. Comp. Phys. {\bf 125}  325--337.

\bibitem{lya} L. V. Yakushevich 2004 {\it Nonlinear Physics of DNA}, Wiley-VCH, Weinheim.

\bibitem{db1} R. Dandoloff and R. Balakrishnan 2005    J. Phys. A: Math.  Gen.  {\bf 38}  6121--6127.







\bibitem{gnt1} A. Goriely, M. Nizette and M. Tabor 2001   J. Nonlin. Sci. {\bf 11}  3--45.





\bibitem{skyrme} J. K. Perring and T. H. R. Skyrme 1962 Nucl. Phys. {\bf 31} 550-555


\bibitem{prl}  R. Dandoloff, S. Villain-Guillot, A. Saxena, and A. R. Bishop 1995 Phys. Rev. Lett. {\bf 74}  813-815.

\end{thebibliography}
\end{document}